\documentclass[a4paper,12pt]{article}
\usepackage{amssymb,amsmath,amsthm}

\newtheorem{proposition}{Proposition}[section]

\theoremstyle{remark}
\newtheorem{remark}{Remark}[section]

\begin{document}

% Definition of title page:
\title{
    Binary black hole spacetimes  with a helical Killing vector
}
\author{C.~Klein,
Max-Planck-Institut f\"ur Mathematik in den Naturwissenschaften, \\
Inselstr. 22,
    04103 Leipzig, Germany
    % insert author(s) here
}
\date{\today}    % optional

\maketitle
\begin{abstract}
Binary black hole spacetimes with a helical Killing vector, which are 
discussed as an approximation for the early stage of a binary system,  are 
studied in a projection formalism. In this setting the four 
dimensional Einstein equations are equivalent to a three dimensional  
gravitational theory with a $SL(2,\mathbb{C})/SO(1,1)$ sigma model as 
the material source. The 
sigma model is determined by a complex  Ernst equation. $2+1$ decompositions
of the 
3-metric are used to establish the field equations on the orbit space of the 
Killing vector. The 
two Killing horizons of spherical topology which characterize the 
black holes, the cylinder of light where the Killing vector changes 
from timelike to spacelike, and infinity are singular points of 
the equations.  The horizon and the 
light cylinder are shown to be regular singularities, i.e.\ the metric 
functions can be expanded in a formal power series in the vicinity. 
The behavior of the metric at spatial infinity is studied in terms of
formal series solutions to the linearized Einstein equations. It is shown that 
the spacetime is not asymptotically flat in the strong sense to have a 
smooth null infinity under the
assumption that the metric tends asymptotically to the Minkowski 
metric.  In this case the metric functions have an oscillatory behavior in the radial coordinate in 
a non-axisymmetric setting, the asymptotic multipoles are not defined. The asymptotic 
behavior of the Weyl tensor near infinity shows that there is no 
smooth null infinity.  
\end{abstract}

\section{Introduction}

Binary black hole systems in the last stage before coalescence are the 
most promising sources of gravitational radiation to be detected with 
the first generation of gravitational wave detectors.
From a theoretical point of view this is a difficult 
relativistic problem which can possibly only be solved numerically since 
there are no symmetries. The advantage of black hole systems is that 
no matter is involved and that only the vacuum equations have to be studied. It 
is generally expected that binary systems will have an early stage of 
quasi-circular motion for large separation of the binaries. Radiation 
damping will lead to almost circular orbits with the radius decreasing 
in time due to the emitted radiation. For smaller distances of the binaries 
this motion is expected to be followed by a rapid inspiral. The end 
result will be a single black hole which will settle to a stationary 
hole in a ring down phase which can be described by black hole 
perturbation theory. For a given binary system an important characteristic quantity
is the innermost stable circular orbit (ISCO), the last almost circular 
orbit before the final inspiral. In general there is no unique 
definition of the ISCO, but it  should be possible to define a characteristic 
scale where the quasi-stationary early phase of the binary system 
comes to an end. 

In the case of the binary motion of two oppositely charged particles,  Sch\"onberg 
\cite{schoenberg} and Schild \cite{schild} considered an approximation 
to the quasi-stationary phase of the system. The 
quasi-circular motion is approximated by a sequence 
of exactly circular orbits which are obtained by exactly compensating the outgoing 
radiation by ingoing radiation. The binding energy of the system as 
defined in \cite{schild}
decreases with the distance of the charges up to some minimal value 
which can be taken as the definition of the ISCO: for smaller values 
of the distance, more and more incoming radiation is needed to 
stabilize the circular motion. The approximation thus predicts its 
own breakdown and allows for an unambiguous definition of the ISCO. 
The quasi-stationary approximation corresponds to a so-called helical 
Killing vector $\xi$ of the system which is in standard Minkowski 
coordinates given by  $\xi=\partial_{t}+\Omega\partial_{\phi}$. The 
main features of such a vector can already be inferred from this case: 
the vector becomes null at the so-called light cylinder given by 
$\rho=1/\Omega$ where the observer rotating with the angular velocity 
$\Omega$ rotates with the velocity of light. In the interior of the 
light cylinder, the Killing vector is timelike, in the exterior 
spacelike. In a general spacetime with a helical Killing vector, the 
light cylinder will be deformed, but will still have cylindrical 
topology. 

Detweiler \cite{detweiler1,blackdet,detweiler2} suggested to 
use this concept to describe the quasi-circular regime of binary 
black holes which corresponds to studying spacetimes with a helical Killing 
vector. Since Einstein's theory is a nonlinear 
theory, the incoming radiation will lead to a  spacetime which is not asymptotically 
flat in a strong sense (mass and angular momentum cannot be defined in 
the usual way asymptotically). It was shown in  \cite{gibstew} that 
spacetimes with a helical Killing vector cannot have a smooth null 
infinity if there is no additional stationary Killing vector close to 
$\mathcal{I}$, see also \cite{ashxan}. Though the 
Arnowitt-Deser-Misner(ADM)-mass cannot be 
defined, Friedman et al. \cite{friedman} could show that a 
thermodynamical treatment as in the single black hole case is 
possible and that there exists a first law. With the help of the 
first law, the ISCO can be defined for asymptotically flat 
spacetimes (which are possible for instance in the case of the first 
post-Newtonian order) as in the Maxwell 
case as the minimum of the binding energy which also marks the onset 
of dynamical instability. It is not clear how this result can be 
generalized to non-asymptotically flat spacetimes in full general 
relativity. 
An important conceptional advantage of spacetimes with a helical Killing 
vector is the fact shown in \cite{friedman} that spatially compact 
Killing horizons are event horizons. This allows for a local 
characterization of the event horizons in these models. It is thus  
not necessary to use local concepts as an apparent horizon or 
the concepts developed in \cite{ashtekarhor}.

In the study of binary black hole system, mainly approximative and 
numerical methods have been used so far. 
Post-Newtonian calculations have been carried out up to the third 
post-Newtonian order including resummation techniques (see 
\cite{buonnano,dis,djs,blanchet,dgg}). Within this approximation the 
determination of the ISCO appears to be self-consistent since corrections 
to its value due to higher order terms can be shown to be negligible. 
The post-Newtonian metric, however, cannot be used close to the 
horizons since black holes are the strongest 
relativistic objects known. Numerical calculations so far have been 
mainly performed within the Isenberg-Wilson-Matthews (IWM) 
theory \cite{IsenbN80,WilsoM89}, an alternative theory of gravitation without 
radiation. It follows from the Einstein equations in a standard 
$3+1$-splitting for a conformally flat spatial metric on the 
$t=const$ hypersurfaces. Only the trace of the 6 time evolution 
equations is considered. By construction the theory coincides with the 
Einstein theory for spacetimes with conformally flat spatial slices
and thus reproduces exactly the first post-Newtonian 
approximation. It has to be noted that the Kerr solution does not 
allow conformally flat spatial hypersurfaces, see \cite{vk1}. In 
\cite{cook94,pfeiffer00,baumgarte00} initial values were constructed 
from the constraint equations via
so called Bowen-York initial data by solving the Lichnerowicz 
equation numerically.  This resulted in a 
significant discrepancy with post-Newtonian results for the ISCO. 
In  \cite{ggb1,ggb2}, complete IWM binary black hole
spacetimes with a helical Killing vector  were constructed 
numerically. The results were in good agreement with post-Newtonian 
results, but suffered from an inconsistency in the model in the form 
of non-regular horizons. This non-regularity appears to be unavoidable 
for IWM spacetimes. To answer the question whether the Einstein equations 
allow for binary black hole spacetimes with smooth 
disconnected horizons in the presence of a helical Killing vector, it
therefore appears necessary to study the fully relativistic 
situation. In \cite{yo} the constraint equations are solved in the 
presence of an approximate helical Killing vector on a background which 
is the superposition of two Kerr-Schild metrics. 

In general relativity, a helical Killing vector will lead to 
Einstein equations which are of a mixed type, i.e.\ elliptic in the 
interior of the light cylinder and hyperbolic in the exterior. Such 
equations are studied in aerodynamics \cite{sonic} in the case of 
transonic flows. For a review on mixed problems arising in 
gravitation, see \cite{stewart}. It was 
shown by Torre \cite{torre} that the resulting equations can be written in the form 
of a first order differential system which belongs to the symmetric 
positive systems of Friedrichs \cite{friedrich}  and Lax and Phillips 
\cite{lax}. Classes of 
boundary conditions compatible with these equations are given in 
\cite{torre}. Numerical studies of two-dimensional equations of this 
type were performed in \cite{whelan1,whelan2}. Three-dimensional toy models 
for the helically reduced Einstein equations were considered 
numerically in \cite{andrade}.

The purpose of this paper is to study the Einstein equations in 
the presence of a helical Killing vector for a vacuum spacetime with 
two disconnected Killing horizons of spherical topology. The idea is 
to derive a set of equations which is well suited for a numerical 
treatment by taking full advantage of the Killing vector. In the vicinity 
of the critical points of the equations as the black hole horizons, 
the metric functions will be given in terms of formal power series 
which should be useful for the numerical implementation.  
The approach is similar as 
in the study of cosmological singularities, see e.g.\ \cite{tod} and 
references therein. 
These methods are purely local and do not imply an existence proof for 
a spacetime with a helical Killing vector and two regular horizons.  
The use of so-called Fuchsian methods (see \cite{rendall}) in the 
context of cosmological singularities would not change the situation 
since even an existence proof for one regular horizon would not imply 
the existence of  the second regular horizon one is interested in here. 
Therefore we will not discuss the question of existence and 
convergence radii of the formal series solutions in this paper, to 
show global 
existence  different methods will have to be applied. Function counting, 
i.e.\ the identification of free functions in the formal series 
solutions indicates, however, that spacetimes with a helical Killing 
vector and two regular horizons could exist: the series in the 
vicinity of the horizon contains two free functions which could be 
fixed in a way to allow for two regular horizons. Such an approach is 
suitable for a numerical treatment which could give -- if 
successful -- a strong indication of the existence for corresponding 
solutions to the Einstein equations.  Though formal expansions do not 
provide existence proofs for solutions, they proved to be very useful and 
reliable to establish the behavior of cosmological solutions near 
singularities and served as a guide to prove existence and 
uniqueness, see \cite{tod,rendall}.

To establish the Einstein equations in the presence of a helical Killing 
vector we use a projection formalism \cite{ehlers,geroch} which 
leads to equations of 3-dimensional gravity with a sigma model as 
material source. The sigma model is determined by a generalized Ernst 
equation. The equations are discussed on the space of orbits of the Killing 
vector in a $2+1$-decomposition of the 3-space. The fixed points of 
the Killing vector, the horizons and the light cylinder where the 
Killing vector changes from being timelike to spacelike are singular points 
of the equations. They will be studied in terms of formal expansions 
of the metric functions in a chosen gauge. In this approach the 
horizons are regular horizons as in the case of the Kerr metric, but 
the series solutions contain two free functions which cannot be fixed 
locally. At the light cylinder the metric shows a similar behavior as 
at an ergosphere.
If the constraints in 
the $2+1$-approach are satisfied by the regularity conditions, the 
only  equations to be solved are the two Ernst equations and the 
three `evolution' (with respect to the radial coordinate) equations. 

The behavior at infinity for spacetimes with a helical Killing vector 
is not yet well understood. There are indications from 
spherically symmetric models with equal 
amounts of ingoing and outgoing null dust \cite{gergely}, which can
be seen as a spherically symmetric analog to a helically symmetric
spacetime, that the spacetime is not even asymptotically flat in the
weak sense that the metric tends asymptotically to the Minkowski
metric. So far  the only rigorous result due to Gibbons 
and Stewart \cite{gibstew} is that there can be no smooth null 
infinity unless there is an additional axial Killing vector. Here we 
show that this is already  case under the possibly too strong 
assumption that the 
spacetime tends to Minkowski spacetime asymptotically. 
Considering the Einstein equations in the presence of a helical 
Killing vector for large values of the radial coordinate, we show  by 
using formal expansions of the metric that the 
spacetime is not asymptotically flat in the strong sense that mass and 
angular momentum can be defined unless there is an additional axial 
Killing vector. The Weyl scalars behave
asymptotically as $1/r$, but the coefficients of the $1/r$ terms have 
an oscillatory dependence on $r$. Thus there is no smooth $\mathcal{I}$ 
and no peeling in accordance with \cite{gibstew}.  In asymptotically
flat spacetimes for a stationary Killing vector, the Komar integral 
can  be used to determine a conserved quantity via a surface integral calculated 
at finite radius, for which then the limit of an infinite radius is 
taken. We show that this limit is not defined under the above 
assumptions unless  there is an additional axial Killing vector. 

The paper is organized as follows: In section 2 we use the 
projection formalism for the vacuum Einstein equations in the 
presence of a helical Killing vector. We discuss the resulting equations 
for the example of Minkowski spacetime. In section 3 we use 
$2+1$-decompositions of the 3-space to study the singularities of the 
equations, the Killing horizons and the light cylinder. We use 
a formal expansion of the metric functions in local coordinates adapted to 
the singularities. In section 4 we study the 
linearized Einstein equations on a Minkowski background 
asymptotically in terms of formal expansions of the metric. 
We discuss the Weyl tensor and the Komar integral. In 
section 5 we add 
some concluding remarks. 

\section{Quotient space metrics and Ernst equations}
The existence of a Killing vector can be used to establish a simplified 
version of the field equations by dividing out the group action. 
These quotient space metrics were first used in \cite{ehlers}, see 
also \cite{geroch}; here we will follow  \cite{maisonehlers}. 
We use adapted coordinates in which the Killing vector $\xi$ is 
given by $\xi=\partial_{t}$ where $t$ is not necessarily a timelike 
coordinate. The norm of the Killing vector will be denoted by $f$. 
The decomposition we are using is not defined at the fixed points of the group 
action, i.e.\ the zeros of $f$, and the resulting equations will be singular 
at the set of zeros of  $f$. The behavior of the solutions at these 
singular points will be discussed in the following section by 
studying formal expansions of the metric in the vicinity 
of the fixed points. 

In contrast to a standard $3+1$-decomposition, the metric is written 
in this approach in the form 
\begin{equation}
    ds^{2}=-f(dt+k_{a}dx^{a})(dt+k_{b}dx^{b})+\frac{1}{f}h_{ab}dx^{a}
    dx^{b}
    \label{maison1};
\end{equation}
latin indices always take the values $1,2,3$ corresponding to the 
spatial coordinate.  
The Einstein equations in vacuum imply a Maxwell-type 
equation for what corresponds to the momentum constraint in a 
standard $3+1$-decomposition (see \cite{maisonehlers})
\begin{equation}
    \frac{1}{2}D_{a}(f^{2}k^{ab})=0
    \label{maison3},
\end{equation}
where $k_{ab}=k_{b,a}-k_{a,b}$ ($D_{a}$ denotes the covariant 
derivative with respect to $h_{ab}$).  Notice that all indices here are raised and 
lowered with $h_{ab}$. If we define the twist potential 
$b$ via ($\epsilon^{abc}$ is the tensor density with 
$\epsilon^{123}=1/\sqrt{h}$)
\begin{equation}
    k^{ab}=\frac{1}{\sqrt{h}f^{2}}\epsilon^{abc}b_{,c}
    \label{maison4},
\end{equation}
where $h$ is the 
determinant of $h_{ab}$, then equation (\ref{maison3}) is identically satisfied.
The potentials $f$ and $b$ can be combined to 
the complex Ernst potential $\mathcal{E}=f+ib$ \cite{ernst}. The 
equations for $f$  and the integrability condition for 
$b$ can then be combined to the generalized
complex Ernst equation (the Ernst equation was originally obtained for 
the stationary axisymmetric case in \cite{ernst})
\begin{equation}
    fD_{a}D^{a}\mathcal{E}=D_{a}\mathcal{E}D^{a}\mathcal{E}
    \label{ein4}.
\end{equation}
The 4 constraint equations in the standard $3+1$-decomposition are thus 
replaced by a single scalar complex equation which is an advantage both for the 
analytical and the numerical treatment. 

The equations for the metric $h_{ab}$ can be written in the form
\begin{equation}
    R_{ab}=\frac{1}{2f^{2}}
    \Re (\mathcal{E}_{,a}\bar{\mathcal{E}}_{,b})
    \label{ein1a},
\end{equation}
where $R_{ab}$ is the three-dimensional Ricci tensor corresponding to 
$h_{ab}$. 
Equations (\ref{ein1a}) describe  three-dimensional gravitation with some 
matter model which turns out to be a $SL(2,\mathbb{C})/SO(1,1)$ sigma 
model, see \cite{sigma}. It is obvious that zeros of the norm of the 
Killing vector are singular points of the equations. 

To illustrate the above equations in the presence of a helical Killing 
vector, it is instructive to consider Minkowski spacetime in a 
rotating frame. In an asymptotically non-rotating frame, the Minkowski 
metric in standard cylindrical coordinates is in the above formalism 
for the stationary Killing vector given by $f=1$, $b=0$ and 
$h_{ab}=\mbox{diag}(1,1,\rho^{2})$. In a rotating coordinate system 
where $\phi'=\phi-\Omega t$ with constant $\Omega$, we get for a 
helical Killing vector $\xi = \partial_{t}+\Omega\partial_{\phi}$ (notice 
that there is a helical Killing vector in Minkowski spacetime for 
arbitrary $\Omega$)
\begin{equation}
    f'=1-\Omega^{2}\rho^{2}, \quad b'=-2\Omega z
    \label{eq:corot},
\end{equation}
where we have put a physically irrelevant constant in the definition 
of the twist potential equal to zero.
Since the spatial metric $h_{ab}$ in (\ref{maison1}) is rescaled by $f$, we have 
$h_{ab}'=f'h_{ab}$ except for $h_{\phi\phi}'$ which is invariant under 
a transformation to a rotating frame. The light cylinder where the rotating observers 
corresponding to the vector $\xi$ move with the velocity of light is given in this case by 
$\rho=1/\Omega$. In the interior of this cylinder, the Killing vector 
is timelike and $f$ is thus positive, in the exterior it is 
spacelike. At the cylinder the signature of the metric $h_{ab}$ 
changes from $+3$ to $-1$. In the four-dimensional picture there is 
no change in the signature of the metric but $t$ and $\phi$ change 
roles at the light cylinder, $\phi$ being a timelike coordinate in the 
exterior of the cylinder. 

The Ernst equation takes in non-rotating 
coordinates the simple form 
\begin{equation}
    f\Delta \mathcal{E}=(\nabla 
\mathcal{E})^{2}
    \label{eq:ernsteq},
\end{equation}
where $\Delta$ and $\nabla$ are the standard differential operators in 
cylindrical coordinates. In the rotating frame, the Laplace 
operator is replaced with the linear operator $\mathcal{L}$ defined by
\begin{equation}
    \mathcal{L}:=\partial_{\rho\rho} +\frac{1}{\rho}\partial_{\rho}+\partial_{zz}
+(1-\Omega^{2}\rho^{2})\frac{1}{\rho^{2}}\partial_{\phi\phi}
    \label{eq:operator},
\end{equation}
which is just the helically reduced flat d'Alembert operator in a 
rotating frame. 
In the axisymmetric case (no $\phi$-dependence) $\mathcal{L}$
reduces to the flat 
Laplace operator. In the non-axisymmetric case solutions to the 
equation $\mathcal{L}\mathcal{E}=0$ behave for small $\rho$ like solutions 
to the Laplace equation
and for large $\rho$ like solutions to a hyperbolic equation. Separating 
the angular dependence in spherical coordinates in a standard way via 
spherical harmonics, 
$\mathcal{E}=\sum_{lm}^{}R_{lm}(r)Y_{lm}(\theta,\phi)$, one recognizes 
that the solutions of $\mathcal{L}\mathcal{E}=0$ behave close to the
origin as $r^{l}$ like solutions 
to the Laplace equation and for $r\to\infty$ as $e^{i\Omega r}/r$. 
One has thus to expect an oscillatory behavior for large $r$. Numerical 
studies of this type of equations have been carried out in 
\cite{whelan1,whelan2,andrade} and \cite{blackdet,detweiler2}. In general, 
Sommerfeld conditions (outgoing wave condition at finite values of 
$r$) have been used. Existence and 
uniqueness of solutions to boundary value problems for these equations 
were studied in \cite{torre} using the theory of symmetric positive 
systems. 

In the stationary axisymmetric case the above equations can be 
further simplified. It can be shown (see e.g.\ \cite{exac}) that 
the spatial metric can be written in this case in the form 
$h_{ab}=\mbox{diag}(e^{2k},e^{2k},\rho^{2})$. In this case the Ernst 
equation decouples from the equations for the metric function $k$ and 
takes the form (\ref{eq:ernsteq}) for a $\phi$-independent Ernst 
potential. The metric function $k$ follows for a given Ernst potential 
in terms of a line integral. In this formalism the Kerr solution for a 
single black hole with mass $m$ and angular momentum 
$J=m^{2}\sin\varphi$ takes a particularly simple form, 
\begin{equation}
    \mathcal{E}  =  \frac{e^{-i\varphi}r_{+}+e^{i\varphi}r_{-}-2m\cos\varphi
    }{e^{-i\varphi}r_{+}+e^{i\varphi}r_{-}+2m\cos\varphi}
    \label{eq:kerr},
\end{equation}
where $r_{\pm}=\sqrt{(z-m\cos\varphi)^{2}+\rho^{2}}$, 
and where the  horizon is given by $\rho=0$ and  $|z|\leq m\cos \varphi$.

Here we consider spacetimes with a single helical Killing vector. 
We adopt the definition of Friedman et al. \cite{friedman} that the 
Killing vector can be written in the form $\xi=\partial_{t'}+\Omega 
\partial_{\phi'}$ where $\partial_{\phi'} $ is a spacelike vector 
with circular orbits of length $2\pi$ unless it vanishes. The vector 
$\partial_{t'}$ is timelike outside the history of some sphere, $\Omega$ 
is a constant, see \cite{friedman} for details. This vector 
generalizes the helical Killing vector of Minkowski spacetime 
discussed above.  It corresponds 
to the introduction of observers corotating with the binary system. 
Close to the black holes this vector will be timelike, but it will 
become null if these observers rotate with the velocity of light 
which determines the so-called light cylinder. 

\section{$2+1$-decomposition, horizons and light cylinder}
In this work we are interested in spacetimes with a helical Killing 
vector that contain binary black holes. 
Since it was shown in \cite{friedman} that spatially compact Killing 
horizons in this case are event horizons, we are interested in 
spacetimes with two disconnected Killing horizons 
of spherical topology. 

With this assumption it seems convenient to 
introduce two systems of spherical coordinates adapted to the horizons 
in a way that one of them  
is given by $r=R=const$. We assume that the spacetime can be globally
foliated by spheres which is not necessary for the analysis below but 
for the planned numerical implementation. Since equations (\ref{ein1a}) describe a model of
three-dimensional gravity, it seems natural to use a $2+1$ decomposition 
of the 3-space with respect to the radial coordinate. 
Let 
$\mathcal{N}_{a}=(\mathcal{A},0,0)$ and 
$\mathcal{N}^{a}=(1,-\mathcal{B}^{\alpha})/\mathcal{A}$ be the unit 
normal to the $r=const$ surfaces; greek indices take the values 2 and 3. Denoting the 
metric of the $r=const$ surfaces with $s_{\alpha\beta}$,
we can write the metric $h_{ab}$ in the form
\begin{equation}
    h_{ab}dx^{a}dx^{b}=s_{\alpha\beta}
    (dx^{\alpha}+\mathcal{B}^{\alpha}dr)
    (dx^{\beta}+\mathcal{B}^{\beta}dr) +\mathcal{A}^{2}dr^{2}
    \label{21.1}.
\end{equation}
The Ricci tensor splits in the standard way (see e.g.\ \cite{mtw}) in 3 
`evolution' equations which contain second derivatives with respect to 
$r$, 
\begin{equation}
    R_{\alpha\beta}=\nabla_{(\alpha}\dot{\mathcal{N}}_{\beta)}+
    2\mathcal{K}_{\alpha\gamma}
    \mathcal{K}^{\gamma}_{\beta}-\dot{\mathcal{N}}_{\alpha}
    \dot{\mathcal{N}}_{\beta}+\pounds_{\mathcal{N}}K_{\alpha\beta}
    +R^{(2)}_{\alpha\beta}-\mathcal{K}_{\alpha\beta}\mathcal{K}
    \label{21.13}
\end{equation}
and 3 `constraint' equations below which have at most first derivatives 
with respect to $r$. Here $\mathcal{K}_{\alpha\beta}$ is the exterior 
curvature of the $r=const$ surfaces given by
\begin{equation}
    \mathcal{K}_{\alpha\beta}=
    \frac{1}{\mathcal{A}}\left(\nabla_{(\alpha}B_{\beta)}-\frac{1}{2}
    s_{\alpha\beta,r}\right)
    \label{21.14},
\end{equation}
where $\nabla_{\alpha}$ denotes the covariant derivative associated 
to $s_{\alpha\beta}$. The Lie derivative of 
$\mathcal{K}_{\alpha\beta}$  in the direction of $\mathcal{N}$ is 
denoted by $\pounds_{\mathcal{N}}K_{\alpha\beta}$, 
\begin{equation}
    \pounds_{\mathcal{N}}\mathcal{K}_{\alpha\beta}
    =\mathcal{K}_{\alpha\beta,r}\frac{1}{\mathcal{A}}
    -\mathcal{K}_{\alpha\beta,\gamma}\frac{\mathcal{B}^{
    \gamma}}{\mathcal{A}}-\mathcal{K}_{\alpha\gamma}\left(
    \frac{\mathcal{B}^{\gamma}}{\mathcal{A}}\right)_{,\beta}
    -\mathcal{K}_{\beta\gamma}\left(
    \frac{\mathcal{B}^{\gamma}}{\mathcal{A}}\right)_{,\alpha}
    \label{21.15},
\end{equation}
and 
$\dot{\mathcal{N}}_{\alpha}=- (\ln\mathcal{A})_{,\alpha}$. 
The two equations corresponding to the momentum 
constraint in the standard $3+1$-decomposition read 
\begin{equation}
    -R_{a\alpha}\mathcal{N}^{a}=\nabla_{\beta}
    \mathcal{K}^{\beta}_{\alpha}-\nabla_{\alpha}
    \mathcal{K}^{\beta}_{\beta}
    \label{21.12},
\end{equation}
the Hamiltonian constraint is given by
\begin{equation}
    R_{ab}\mathcal{N}^{a}\mathcal{N}^{b}-
    s^{\alpha\beta}R_{\alpha\beta}=\mathcal{K}^{2}
    -\mathcal{K}_{\alpha\beta}\mathcal{K}^{\alpha\beta}-R^{(2)}.
    \label{21.11a}
\end{equation}
where $\mathcal{K}=\mathcal{K}^{\alpha}_{\alpha}$ and 
$R^{(2)}=s^{\alpha\beta}R^{(2)}_{\alpha\beta}$. 
In the above equations, greek indices are raised and 
lowered with $s_{\alpha\beta}$ and its inverse.

For concrete calculations one has to fix a gauge. 
Since we are considering horizons of spherical topology and an 
adapted coordinate system, a natural choice of the metric 
$s_{\alpha\beta}$ would be the standard metric 
of the 2-sphere. This is in general possible for smooth metrics which 
we are interested in here since we are looking for regular horizons. 
Due to the fact that the spatial metric $h_{ab}$ is rescaled by the 
norm of the Killing vector which vanishes at the horizon, the metric 
$h_{ab}$ is expected to vanish there as well. A possible choice would 
then be a metric conformal to the metric of the 2-sphere, 
\begin{equation}
    s_{\alpha\beta}=fr^{2}\mbox{diag}(1,\sin^{2}\theta)
    \label{eq:2sphere}.
\end{equation}
This gauge will be convenient in the vicinity of the horizon. As the 
considerations for Minkowski spacetime in the previous section have 
shown, this choice is not possible across the light cylinder because 
of the signature change of the spatial metric there. Since we are 
interested in setting up equations suited for numerical treatment, we 
are looking for a system of coordinates which is able to cover the 
whole spacetime in the exterior of the horizon with a single coordinate 
patch. This could be possible with coordinates in which the angular 
part $g_{\alpha\beta}$ of the four-dimensional metric is the standard 
metric of the 2-sphere as in  Schwarzschild coordinates. The 
latter would imply however the explicit inclusion of the vector $k_{a}$ 
in the 
equations which is contrary to the philosophy of the present approach 
to work only with its dual, the scalar twist potential $b$. 

Therefore we choose here a generalized Weyl gauge which we call 
`quasi-isotropic'. We write 
\begin{equation}
    s_{\alpha\beta}=\mbox{diag}(r^{2}\mathcal{A}^{2},\mathcal{C})
    \label{eq:quasi}.
\end{equation}
A possible choice for $\mathcal{C}$ is $\mathcal{C}=r^{2}\sin^{2}\theta 
(1-R^{2}/r^{2})^{2}$. This corresponds to standard Weyl coordinates 
in which the horizon of a Kerr black hole is a sphere of radius $R$. 
These coordinates are related to the coordinates (\ref{eq:kerr}) via 
$\rho = (r-R^{2}/r)\sin\theta$ and $z=(r+R^{2}/r)\cos\theta$. 
The quasi-isotropic gauge thus reduces to the standard Weyl coordinates 
in the 
axisymmetric case. It can be used in principle throughout the light cylinder, 
but it 
remains to be shown whether it can be used globally.  Due to the 
divergence structure of the Ernst equation (\ref{ein4}) which can be 
written free of covariant derivatives in the form 
\begin{equation}
    f(h^{ab}\sqrt{h}\mathcal{E}_{,a})_{,b}=h^{ab}\sqrt{h}\mathcal{E}_{,a}
\mathcal{E}_{,b}
    \label{eq:ernst},
\end{equation}
the equation has in this gauge the standard terms of the Laplace operator 
for the $rr$ and $\theta\theta$ derivatives. This helps in the 
numerical treatment of the equations since standardized  differential 
operators can be numerically inverted. In this gauge we have the non-vanishing 
Christoffel symbols corresponding to $s_{\alpha\beta}$
\begin{equation}
    \Gamma^{2}_{22}=(\ln \mathcal{A})_{,\theta},
    \Gamma^{2}_{23}=(\ln \mathcal{A})_{,\phi},
    \Gamma^{2}_{33}=-\frac{1}{\mathcal{A}^{2}}(1-R^{2}/r^{2})^{2}\sin\theta 
    \cos\theta,
    \label{bl3}
\end{equation}
and 
\begin{equation}
    \Gamma^{3}_{22}=-\frac{\mathcal{A}\mathcal{A}_{,\phi}}{(1-R^{2}/r^{2})^{2} \sin^{2}\theta},
	 \Gamma^{3}_{23}=\cot\theta.
    \label{eq:bl3a}
\end{equation}
The components of the Ricci tensor read $R^{(2)}_{23}=0$ and
\begin{equation}
    -R^{(2)}_{22}= 
    -\frac{\mathcal{A}^{2}}{(1-R^{2}/r^{2})^{2}\sin^{2}\theta}R^{(2)}_{33}=
    \frac{\mathcal{A}\mathcal{A}_{,\phi\phi}}{(1-R^{2}/r^{2})^{2}\sin^{2}\theta}-1-
    \cot \theta (\ln \mathcal{A})_{,\theta}
    \label{bl4}.
\end{equation}

The 
equations for $R_{ab}$ can be treated as in the case of a $3+1$-decomposition: if 
the constraints are satisfied for some value of $r$, this will be the 
case for solutions to the evolution equations for all values of $r$. 
Since the horizon is a singularity for the equations, one has to give 
boundary conditions there which are compatible with the constraints 
and the evolution equations. With these boundary conditions one has to solve the Ernst equation 
which corresponds to two real equations and the three evolution 
equations (\ref{21.13})\footnote{As in \cite{constraint}, a fully 
constraint approach can be used alternatively in the sense that only the constraint 
equations instead of the evolution equations are solved.}. 
Thus one has to solve in total 5 equations 
as in the case of the IWM problem in \cite{ggb2}. The difference is 
here that the equations are not elliptic in the exterior of the light 
cylinder in contrast to the IWM equations and that the spacetime will 
not be asymptotically flat as discussed in the following section. 

To study the behavior of the metric at the horizon, we use formal 
power series in the local coordinate $y=r-R$. As in the case of ordinary 
differential equations, we adopt for a function $F(r,\theta,\phi)$ the 
ansatz 
\begin{equation}
    F(r,\theta,\phi)=y^{n_{F}}\sum_{j=0}^{\infty}F_{j}(\theta,\phi) y^{j}
    \label{eq:ansatz},
\end{equation}
but here with coefficients $F_{j}(\theta,\phi)$ depending on $\theta$ 
and $\phi$. The question is whether there are formal solutions to the 
Einstein equations of this form with vanishing norm $f$ of the Killing vector for 
$y=0$ which are 
more general than the Kerr solution for a single black hole. Here we 
are only interested in providing formal solutions intended for the 
use in the numerical treatment. Therefore we do neither discuss the 
convergence of the series nor global questions. We get 
(again we ignore a physically irrelevant constant in the definition of $b$)\\
\begin{proposition}\label{3.1}
    In the gauge (\ref{eq:quasi}), the equations (\ref{ein4}) and 
    (\ref{ein1a}) have formal solutions of the form (\ref{eq:ansatz}) with 
    \begin{equation}
        f=f_{0}(\theta,\phi)y^{2}+f_{1}(\theta,\phi)y^{3}+f_{2}(\theta,\phi)y^{4}+\ldots, \quad 
        b=b_{0}(\theta,\phi)y^{4}+b_{1}(\theta,\phi)y^{5}+b_{2}(\theta,\phi)y^{6}+\ldots
        \label{eq:horex},
    \end{equation}
    and 
    \begin{equation}
        \mathcal{A}=\mathcal{A}_{0}(\theta,\phi)y 
        +\mathcal{A}_{1}(\theta,\phi)y^{2}+\mathcal{A}_{2}(\theta,\phi)y^{3}+\ldots,
        \quad \mathcal{B}_{2}=\Theta_{0}(\theta,\phi)y^{3}+\ldots, \quad 
        \mathcal{B}_{3}=\Phi_{0}(\theta,\phi)y^{3}+\ldots
        \label{eq:solhor}.
    \end{equation}
    The functions $f_{0}(\theta,\phi)$ and $b_{0}(\theta,\phi)$ are free 
    functions of $\theta$ and $\phi$ with $f_{0,\phi}(0,\phi)=0$. All other coefficient 
    functions in the expansions (\ref{eq:horex}) and (\ref{eq:solhor}) 
    can be expressed in dependence of $f_{0}(\theta,\phi)$ and 
    $b_{0}(\theta,\phi)$, the leading order terms being
    \begin{equation}
        \mathcal{A}_{0}(\theta,\phi)=\kappa f_{0}(\theta,\phi), \quad 
        \frac{\mathcal{A}_{1}}{\mathcal{A}_{0}}=-\frac{3}{2R},\quad 
        f_{1}=-\frac{f_{0}}{R}, \quad b_{1}=-\frac{2b_{0}}{R},
        \label{eq:a01}
    \end{equation}
    where the constant $\kappa$ is given by $\kappa = 2/(Rf_{0}(0,0))$. 
\end{proposition}

The constant $\kappa$ in the relation between $\mathcal{A}_{0}$ and $f_{0}$ 
indicates a freedom in the choice of $f_{0}(0,0)$. This freedom is 
due to the fact that a scale in the norm of the Killing vector is not 
fixed, after multiplication with some constant, $\xi$ is still a 
Killing vector.

\emph{Proof:}\\
With (\ref{eq:horex}) and (\ref{eq:solhor}), we get for the exterior 
curvature (\ref{21.14}) by using (\ref{bl3}) and (\ref{eq:bl3a})
\begin{eqnarray}
    \mathcal{K}_{22} & = & -\mathcal{A}_{0}(R^{2}+3Ry+2y^{2})-
    \mathcal{A}_{1}(2R^{2}y+5Ry^{2})-3R^{2}\mathcal{A}_{2}y^{2}\nonumber\\
    &&+(\Theta_{0,\theta}-(\ln 
    \mathcal{A}_{0})_{,\theta}\Theta_{0})\frac{y^{2}}{\mathcal{A}_{0}} 
    +\frac{R^{2}\mathcal{A}_{0,\phi}\Phi_{0}}{4\sin^{2}\theta}y^{2}+0(y^{3})
    \nonumber , \\
    \mathcal{K}_{23} & = & \left(\frac{1}{2}\Theta_{0,\phi}-(\ln 
    \mathcal{A}_{0})_{,\phi}\Theta_{0}\right)\frac{y^{2}}{\mathcal{A}_{0}}+
    \left(\frac{1}{2}\Phi_{0,\theta}-\cot\theta \Phi_{0}\right)
    \frac{y^{2}}{\mathcal{A}_{0}}+0(y^{3})
    \nonumber,  \\
    \mathcal{K}_{33} & = & 
    -\frac{4}{\mathcal{A}_{0}}\sin^{2}\theta\left(1-\frac{3}{2R}y+\frac{5y^{2}}{2R^{2}}
    -\frac{\mathcal{A}_{1}}{\mathcal{A}_{0}}y\left(1-\frac{3y}{2R}\right)+\frac{\mathcal{A}_{1}^{2}}{
    \mathcal{A}_{0}^{2}}y^{2}-\frac{\mathcal{A}_{2}}{\mathcal{A}_{0}}y^{2}
    \right)\nonumber\\
    &&+\Phi_{0,\phi}\frac{y^{2}}{\mathcal{A}_{0}} 
    +\frac{4\sin\theta\cos\theta}{R^{2}\mathcal{A}_{0}^{3}}\Theta_{0}y^{2}+0(y^{3}).
    \label{eq:horn1}
\end{eqnarray}
It is straight forward to check that the Hamiltonian constraint 
(\ref{21.11a}) is satisfied to leading order (which is $1/y^{4}$). The momentum 
constraint (\ref{21.12}) leads  in lowest order to
\begin{equation}
    (\ln \mathcal{A}_{0})_{,\alpha}=(\ln f_{0})_{,\alpha}
    \label{eq:mom1}.
\end{equation}
Thus we have $\mathcal{A}_{0}=\kappa f_{0}$ with $\kappa=const$. 

To ensure a regular axis ($\theta=0$) in spherical coordinates, the 
axis must be `elementary flat', i.e.\ small circles around the axis 
must have an invariant circumference of $2\pi$ times the invariant 
radius in the limit of vanishing radius. This means for $\rho=r\sin\theta$, $z=r\cos\theta$
\begin{equation}
    \lim_{\rho\to0}\int_{0}^{2\pi}\sqrt{g_{\phi\phi}}d\phi=2\pi 
    \lim_{\rho\to0}\int_{0}^{\rho}\sqrt{g_{\rho\rho}(\rho',z,\phi)}d\rho' .
\end{equation}
With the above relations this implies ($k_{a}$ is bounded at the 
horizon) 
\begin{equation}
    r\sin\theta\left(1-\frac{R^{2}}{r^{2}}\right) 
    \frac{1}{2\pi}\int_{0}^{2\pi}\frac{1}{\sqrt{f(r,0,\phi)}}d\phi = 
    \frac{\mathcal{A}(r,0,\phi)}{\sqrt{f(r,0,\phi)}}
    \label{eq:axisreg1}.
\end{equation}
Expanding this relation in $y$ and using (\ref{eq:horex}), 
(\ref{eq:solhor}) and (\ref{eq:mom1}), we get in lowest order of $y$ 
the condition that $f_{0}(0,\phi)$ must be independent of $\phi$ and 
thus be a constant. This constant is related to $\kappa$ via 
\begin{equation}
    \kappa= \frac{2}{Rf_{0}(0,0)}
    \label{eq:axisreg2}.
\end{equation}

The Ernst equation (\ref{ein4}) is satisfied in leading order for 
arbitrary $f_{0}(\theta,\phi)$ and $b_{0}(\theta,\phi)$, in the next 
higher order the real part implies
\begin{equation}
    f_{1}+\frac{f_{0}}{R}=0
    \label{eq:ernst1a},
\end{equation}
whereas the imaginary part leads to 
\begin{equation}
    -8b_{0}f_{1}+\frac{2}{R}b_{0}f_{0}+5b_{1}f_{0}=0
    \label{eq:ernst1b}.
\end{equation}
This determines $f_{1}$ and $b_{1}$ in dependence of $f_{0}$ and $b_{0}$. 
In order $y^{n}$, the leading terms in the real part of the Ernst equation 
are 
\begin{equation}
    (n-2)^{2}f_{0}f_{n-2}
    \label{eq:ernstnreal}
\end{equation} 
and 
\begin{equation}
    n(n-4)f_{0}b_{n-4}+8(2-n)b_{0}f_{n-2}
    \label{eq:ernstnim}
\end{equation}
for the imaginary part. 
The Ernst equation can thus be used to all orders determine $f_{n-2}$ and 
$b_{n-4}$ 
in dependence of quantities of lower order. 

With (\ref{eq:horn1}) we get for the Lie derivative of the exterior 
curvature (\ref{21.15}) 
\begin{eqnarray}
    \pounds_{\mathcal{N}}\mathcal{K}_{22} & = & 
    -\frac{1}{y}\left(3R+2R^{2}\frac{\mathcal{A}_{1}}{\mathcal{A}_{0}}\right)-4-
    7R\frac{\mathcal{A}_{1}}{\mathcal{A}_{0}}
    +2R^{2}\frac{\mathcal{A}_{1}^{2}}{\mathcal{A}_{0}^{2}}-6R^{2}\frac{\mathcal{A}_{2}}{\mathcal{A}_{0}}
    \nonumber\\
    &&+\frac{4\Theta_{0,\theta}}{\mathcal{A}_{0}^{2}}-7(\ln 
    \mathcal{A}_{0})_{,\theta}\frac{\Theta_{0}}{\mathcal{A}_{0}^{2}}+\frac{3R^{2}(\ln \mathcal{A}_{0})_{,\phi}
    \Phi_{0}}{4\sin^{2}\theta}+0(y)
    \nonumber  \\
    \pounds_{\mathcal{N}}\mathcal{K}_{23} & = &  \frac{1}{\mathcal{A}_{0}^{2}}\left(2
    \Theta_{0,\phi}-5(\ln \mathcal{A}_{0})_{,\phi}\Theta_{0}+2\Phi_{0,\theta}
    -4\cot\theta\Phi_{0} -(\ln \mathcal{A}_{0})_{,\theta}\Phi_{0}\right)+0(y)
    \nonumber  \\
    \pounds_{\mathcal{N}}\mathcal{K}_{33} & = & 
    \frac{4\sin^{2}\theta}{\mathcal{A}_{0}^{2}y}\left(\frac{3}{2R}+
    \frac{\mathcal{A}_{1}}{\mathcal{A}_{0}}\right)-\frac{4\sin^{2}\theta}{\mathcal{A}_{0}^{2}}\left(
    \frac{5}{R^{2}}+\frac{9\mathcal{A}_{1}}{2R\mathcal{A}_{0}}+\frac{3\mathcal{A}_{1}^{2}}{\mathcal{A}_{0}^{2}} 
    -\frac{2\mathcal{A}_{2}}{\mathcal{A}_{0}}\right)\label{eq:horn2}\\
    &&+16\sin\theta\cos\theta\frac{\Theta_{0}}{R^{2}\mathcal{A}_{0}^{4}}-
    4\sin^{2}\theta(\ln \mathcal{A}_{0})_{,\theta}
    \frac{\Theta_{0}}{R^{2}\mathcal{A}_{0}^{4}}+
    \frac{4\Phi_{0,\phi}}{\mathcal{A}_{0}^{2}}-3(\ln 
    \mathcal{A}_{0})_{,\phi}\frac{\Phi_{0}}{\mathcal{A}_{0}^{2}}
    +0(y)
    \nonumber
\end{eqnarray}
Since there are second order derivatives with respect to $r$ in the 
the evolution equations as in the Ernst equation, higher order terms in the 
expansion of the metric functions will appear here before they do in 
the constraints. Therefore there are no further 
conditions on the lowest order terms. In order $1/y$ the equation for 
$R_{22}$ (there are no 
contributions from the Ernst potential in this order) leads to 
\begin{equation}
    \frac{R^{2}}{y}\left(\frac{3}{2R}+\frac{\mathcal{A}_{1}}{\mathcal{A}_{0}}\right) = 0
    \label{eq:A1}.
\end{equation}
This implies that there are no terms of order $y$ in the exterior 
curvature (\ref{eq:horn1}) and no terms of  order $1/y$ in the 
Lie-derivatives (\ref{eq:horn2}). 

The leading terms in order $y^{n-3}$ in the evolution equations are 
for $n>2$ in $R_{22}$
\begin{eqnarray}
    &&-R^{2}(1-n)^{2}\frac{\mathcal{A}_{n-1}}{\mathcal{A}_{0}}
     +\frac{1}{\mathcal{A}_{0}^{2}}\left(n\Theta_{n-3,\theta}-(n+3)(\ln 
     \mathcal{A}_{0})_{,\theta}\Theta_{n-3}+\cot\theta 
     \Theta_{n-3}\right)\nonumber  \\
     &  & +\frac{R^{2}}{4\sin^{2}\theta 
     }\left(\Phi_{n-3,\phi}+(n-1)
     (\ln \mathcal{A}_{0})_{,\phi}\Phi_{n-3}\right)
    \label{eq:r22}.
\end{eqnarray}
Similarly we get for $R_{23}$
\begin{equation}
     \frac{1}{2\mathcal{A}_{0}^{2}}\left((n-1)\Theta_{n-3,\phi}-2n(\ln 
     \mathcal{A}_{0})_{,\phi}\Theta_{n-3}+(n-1)
     \Phi_{n-3,\theta}-2(n-1)\cot\theta\Phi_{n-3}-2(\ln 
     \mathcal{A}_{0})_{,\theta}\Phi_{n-3}\right)
    \label{eq:r23},
\end{equation}
and for $R_{33}$
\begin{eqnarray}
    &  & \frac{4\sin^{2}\theta}{R^{2}\mathcal{A}_{0}^{4}}\left(\Theta_{n-3,\theta}-2(\ln 
     \mathcal{A}_{0})_{,\theta}\Theta_{n-3}+n\cot\theta\Theta_{n-3}\right)\nonumber\\
     &&
     +\frac{1}{\mathcal{A}_{0}^{2}}\left(
     n\Phi_{n-3,\phi}-2(\ln 
     \mathcal{A}_{0})_{,\phi}\Phi_{n-3}\right)
    \label{eq:r33}.
\end{eqnarray}
It is straight forward to solve the equations \ref{eq:r23})  and 
(\ref{eq:r33}) for $\Theta_{n-3}$ and $\Phi_{n-3}$.  Function 
$\mathcal{A}_{n-1}$ then follows from equation (\ref{eq:r22}). The 
Ernst equations (\ref{eq:ernstnreal}) and (\ref{eq:ernstnim}) 
determine consequently
$f_{n-2}$ and $b_{n-4}$.  We have thus shown that the evolution 
equations and the Ernst equation can be solved in this way to all orders. 
This completes the proof. 

The fact that $\mathcal{A}_{n-1}$ does not appear in the equations 
(\ref{eq:r23}) and (\ref{eq:r33}) implies that $\Theta$ and $\Phi$ 
can be chosen to vanish for $\phi$-independent $f_{0}$ and $b_{0}$. This leads as expected to the 
Kerr solution. Note that $f_{n-2}$, $b_{n-4}$ and $\mathcal{A}_{n-1}$ 
are determined algebraically by the above equations. Just to 
determine $\Theta_{n-1}$ and $\Phi_{n-1}$, one has to integrate which 
leads to free integration functions. The latter are related to the fact 
that the used gauge conditions do not fix the gauge completely. 
There are transformations of the form 
\begin{equation}
    r' = r+y^{3}P(\theta,\phi)+\ldots,\quad \theta' = 
    \theta+y^{2}S(\theta,\phi)+\ldots,\quad 
    \phi'=\phi+y^{2}T(\theta,\phi)+\ldots
    \label{eq:gauge}
\end{equation}
for non-trivial $P$, $S$ and $T$ which do not change the gauge. Since 
$h_{r\theta}'=h_{r\theta}+h_{\theta\theta}2yS$ and similarly for $h_{r\phi}$, 
there are gauge modes in $\Theta$ and $\Phi$ which show up in the form 
of free integration functions.

\begin{remark}\label{2.1}
    Due to the homogeneity of the Ernst 
    equation in the Ernst potential,  the functions $f_{0}(\theta,\phi)$ and $b_{0}(\theta,\phi)$ are 
    not determined in the above expansions in the vicinity of a horizon. 
    This gives hope that there might be a second regular horizon of 
    spherical topology in the spacetime for a suitable choice of these functions.
    Whereas the 
    behavior of the Ernst potential with respect to $y$ is the same as in 
    the case of a single Kerr black hole, the functions $f_{0}$ and 
    $b_{0}$ may be different. 
\end{remark}

To treat the light cylinder we use a similar approach as for the 
horizon. By the definition of the Killing vector of \cite{friedman} we 
are using here, the light cylinder  will have cylindrical topology. We 
assume that the spacetime can be foliated by cylindrical surfaces, and 
use cylindrical coordinates in which the light  cylinder 
is given by $\rho=\rho_{0}=const$. In an abuse of notation we use the 
same symbols for the $2+1$-decomposition as used for the spherical 
case, 
\begin{equation}
    h_{ab}dx^{a}dx^{b}=s_{\alpha\beta}
    (dx^{\alpha}+\mathcal{B}^{\alpha}d\rho)
    (dx^{\beta}+\mathcal{B}^{\beta}d\rho) +\mathcal{A}^{2}dr^{2}
    \label{zyl1}.
\end{equation}
where $dx^{2}=dz$. We use again the quasi-isotropic gauge which reads 
in this case
\begin{equation}
    s_{\alpha\beta}=\mbox{diag}(\mathcal{A}^{2},\mathcal{C})
    \label{eq:quasizyl}.
\end{equation}
The choice $\mathcal{C}= \rho^{2}$ is possible near the cylinder.

We assume that $f$ has a zero of 
first order in $v=\rho-\rho_{0}$ at the cylinder since the Killing 
vector is supposed to change from timelike to spacelike there.  Again 
we consider formal expansions of the metric function of the form 
\begin{equation}
    F(\rho,z,\phi)=v^{n_{F}}\sum_{j=0}^{\infty}F_{j}(z,\phi) v^{j}
    \label{eq:ansatzzyl},
\end{equation}
We get \\
\begin{proposition}
    In the gauge (\ref{eq:quasizyl}) the equations (\ref{ein4}) and 
    (\ref{ein1a}) have formal power series solutions of the form (\ref{eq:ansatzzyl}) with
    \begin{eqnarray}
        f & = & f_{0}(z)v+f_{1}(z,\phi)v^{2}+\ldots,
    \quad    b  =  b_{0}(z)+b_{2}(z,\phi)v^{2}+\ldots,
        \nonumber  \\
        \mathcal{A} & = & 
        \mathcal{A}_{0}(z)\sqrt{v}+\mathcal{A}_{1}(z,\phi)v^{\frac{3}{2}}+\ldots
    \quad 
    \mathcal{B}_{2}  = Z_{0}(z,\phi)v^{2}+\ldots, \quad 
    \mathcal{B}_{3}=\Phi_{0}(z,\phi)v
        \label{eq:light1}.
    \end{eqnarray}
    The functions $b_{0}(z)$, $\mathcal{A}_{0}(z)$  and $f_{2}(z,\phi)$ are free 
    functions of $z$ and $z$, $\phi$ respectively.  All other coefficient 
    functions in the expansion (\ref{eq:light1}) can be expressed in 
    dependence of $b_{0}$, $\mathcal{A}_{0}$ and $f_{2}$,
    the leading term being
    \begin{equation}
        b_{0,z}^{2}=f_{0}^{2}
        \label{eq:light1a}.
    \end{equation}
\end{proposition}

Proof:\\
The formulas for the $2+1$-decomposition in section 2 apply with the 
trivial change that $r$ has to be replaced by $\rho$. The chosen gauge 
is also very similar to the one used for the horizon, the only 
difference being the factor $r^{2}$ in the expression for $s_{22}$. 
Therefore we will not give explicit formulas for the Christoffel 
symbols and $R_{\alpha\beta}^{(2)}$ here. For the exterior curvature 
(\ref{21.14}), we get with (\ref{eq:light1})
\begin{eqnarray}
    \mathcal{K}_{22} & = & 
    -\frac{\mathcal{A}_{0}}{2\sqrt{v}}-\frac{3}{2}\mathcal{A}_{1}\sqrt{v}+0(v^{\frac{3}{2}}),
    \nonumber  \\
    \mathcal{K}_{23} & = & 
    \frac{\sqrt{v}}{2\mathcal{A}_{0}}\Phi_{0,z}+0(v^{\frac{3}{2}}),
    \nonumber  \\
    \mathcal{K}_{33} & = & 
    -\frac{\rho}{\mathcal{A}_{0}\sqrt{v}}\left(1-\frac{\mathcal{A}_{1}}{\mathcal{A}_{0}}v\right)+
    \frac{\sqrt{v}}{\mathcal{A}_{0}}\Phi_{0,\phi}+0(v^{\frac{3}{2}})
    \label{eq:lightn1}.
\end{eqnarray}
This implies for the Lie derivative of the exterior curvature
\begin{eqnarray}
    \pounds_{\mathcal{N}}\mathcal{K}_{22} & = & 
    \frac{1}{4v^{2}}-\frac{\mathcal{A}_{1}}{\mathcal{A}_{0}v} 
    +0(v^{0}),
    \nonumber  \\
    \pounds_{\mathcal{N}}\mathcal{K}_{23} & = & 
    \frac{1}{4\mathcal{A}_{0}^{2}v}\Phi_{0,z}+0(v^{0}),
    \nonumber  \\
    \pounds_{\mathcal{N}}\mathcal{K}_{33}  & = & 
    \frac{\rho_{0}}{2\mathcal{A}_{0}^{2}v^{2}}-\frac{1}{2\mathcal{A}_{0}^{2}v}
    +\frac{\Phi_{0,\phi}}{2\mathcal{A}_{0}^{2}v}
    +0(v^{0})
    \label{eq:lightn2}.
\end{eqnarray}

In lowest order, the equation for $R_{22}$ then yields 
(\ref{eq:light1a}),
whereas the relations for $R_{23}$ and $R_{33}$ can only be satisfied 
in this case for
\begin{equation}
    b_{0,\phi}=0
    \label{eq:lightn3}.
\end{equation}
The $z$-component of the momentum constraint implies 
\begin{equation}
    2b_{0,z}b_{2}+f_{0,z}f_{0}-\frac{Z_{0}}{\mathcal{A}_{0}^{2}}b_{0,z}^{2}=0
    \label{eq:light3},
\end{equation}
which determines $b_{2}$, 
whereas  the $\phi$-component of the momentum constraint gives 
\begin{equation}
    \mathcal{A}_{0,\phi}=0
    \label{eq:light3a}.
\end{equation}
With condition (\ref{eq:light1a}) 
the Hamiltonian constraint is satisfied to leading order. 

The real 
part of the Ernst equation gives no additional condition in leading order, in order 
$v$ it leads to 
\begin{equation}
    f_{1}=\frac{f_{0}}{2\rho_{0}}
    \label{eq:lightn4}.
\end{equation}
The imaginary part gives with (\ref{eq:light1a}) in leading order 
(\ref{eq:light3a}). Consequently $b_{2}$ and $f_{1}$
are determined in this order.

The equation for $R_{22}$ reads with (\ref{21.13}) in order $1/v$ 
\begin{equation}
    \frac{\mathcal{A}_{1}}{\mathcal{A}_{0}}=
    \frac{1}{\rho_{0}}
    \label{eq:lightn6}.
\end{equation}
In the same order  $R_{23}$ is identically satisfied.  $R_{33}$ takes 
the form
\begin{equation}
    0=\frac{2\mathcal{A}_{1}}{\mathcal{A}_{0}} 
    +\frac{1}{\rho_{0}}\Phi_{0,\phi}
    \label{eq:lightn8}.
\end{equation}
This fixes $\Phi_{0}$ and $\mathcal{A}_{1}$. Thus the functions 
$b_{0}(z)$ and 
$\mathcal{A}_{0}(z)$ are not determined by the above equations.  In 
addition $Z_{0}$ is not yet fixed.

In higher orders of the expansion, the reasoning will be similar since 
the general structure of the equations is the same: the 
Ernst equation in order $v^{n-1}$ with $n>1$ contains the leading 
terms $n(n-3)f_{0}f_{n-1}$ and 
\begin{equation}
    n(n-3)b_{n}+2b_{0,z}\frac{Z_{n-2}}{\mathcal{A}^{2}_{0}}
    \label{eq:lightn9}
\end{equation}
in the real and the imaginary part respectively.  
The evolution equations in order $v^{n-2}$ contain the leading terms
\begin{equation}
    -\frac{b_{n,z}}{b_{0,z}}-\left(n^{2}-n-\frac{1}{4}\right) \frac{\mathcal{A}_{n}}{
    \mathcal{A}_{0}}+\frac{1}{\mathcal{A}_{0}^{2}}\left(\left(n-\frac{1}{2}\right)Z_{n-2,z}-
    (n-2)(\ln \mathcal{A}_{0})_{,z}Z_{n-2}\right)
    \label{eq:22n},
\end{equation}
\begin{equation}
    -\mathcal{A}_{0}^{2}\frac{b_{n,\phi}}{b_{0,z}}+nZ_{n-1,\phi}+\left(n-\frac{1}{2}\right)\Phi_{n-1,z}
    \label{eq:23n},
\end{equation}
and 
\begin{equation}
    (2n-1)\frac{\mathcal{A}_{n}}{\mathcal{A}_{0}}+\frac{n}{2\rho_{0}}\Phi_{n-1,\phi}
    -\frac{1}{\mathcal{A}_{0}^{2}}
    Z_{n-2,z}
    \label{eq:33n}.
\end{equation}
Thus one can  determine $f_{n-1}$,
$b_{n}$, $\mathcal{A}_{n}$, $Z_{n-1}$ and $\Phi_{n-1}$ from the above 
equations unless $n=3$. In this case the real part of the Ernst equation 
determines the function $Z_{0}$ which was still free, the function 
$f_{2}$ remains undetermined. 
The equations in higher order fix
all expansion functions except $\mathcal{A}_{0}$,  $b_{0}$ and 
$f_{2}$. Free functions  in $Z$ and $\Phi$ occurring after integration are again 
related to residual gauge freedoms as was the case in the vicinity of 
the horizon. Thus the series 
(\ref{eq:light1}) provide a formal solution in the vicinity of the light cylinder. This 
completes the proof.

\begin{remark}\label{4.3}
    It should be possible to apply Fuchsian methods as in 
    \cite{rendall} to prove existence of the above solutions near the 
    horizon and the light cylinder for some non-vanishing radius of 
    convergence. However this would not answer the 
    decisive question whether there can be two smooth horizons and a 
    smooth light cylinder in the spacetime. Therefore we will not apply 
    these methods here. 
\end{remark}

\section{Asymptotic behavior}
The formal solutions in terms of a series in the vicinity of the two 
horizons and the light cylinder in the previous section obviously do 
not imply global existence of a solution describing a spacetime with 
a helical Killing vector and two regular Killing horizons. The radius 
of convergence of these series is unknown. Therefore it is also not 
possible to make precise statements on the asymptotic behavior of the 
metric. As shown in \cite{gibstew} such spacetimes cannot be 
asymptotically flat in the strong sense that they have a smooth null 
infinity. However this does not exclude the possibility that the spacetime 
is weakly asymptotically 
flat in the sense that the spacetime tends to the Minkowski spacetime. 
This is what we will show in this section though the assumption of an 
asymptotic Minkowski metric might be too strong as indicated by the
work of  \cite{gergely}. However the used techniques 
will also be applicable for even weaker asymptotic conditions.

Since the coordinate system we were using in the previous sections is 
asymptotically rotating (see the considerations for Minkowski 
spacetime in section 2), the Ernst potential is expected to have the 
kinematic terms (\ref{eq:corot}). These terms  will lead to technical 
difficulties if one wants to consider an expansion of the metric 
functions in powers of $1/r$. Therefore we will consider in this 
section asymptotically non-rotating coordinates $(t',r,\theta,\phi')$ 
where $\phi'=\phi+\Omega t$, $t'=t$. The metric will be 
studied via the linearized Einstein equations on a Minkowski 
background. Due to the helical symmetry the metric functions depend on 
$t'$ and $\phi'$ only via the combination $x=\phi'-\Omega t'$. 

We assume the metric to be of the form 
$g_{AB}=\eta_{AB}+\delta_{AB}$ (capital indices take the values 
0,1,2,3) for $r\to \infty$ where 
$\eta_{AB}=\mbox{diag}(-1,1,r^{2},r^{2}\sin^{2}\theta)$ is the 
Minkowski metric in spherical coordinates, and where $\delta_{AB}$ 
gives the deviation from Minkowski spacetime 
for large $r$. In cartesian coordinates 
$\delta_{AB}$ is assumed to be of order $1/r$. An analysis of the 
equations as in section 2 indicates however that the terms of order 
$1/r$ will have an oscillatory dependence in $r$ (we exclude here 
possible logarithmic terms in the metric function).  We will therefore 
consider a formal expansion of the metric functions of the form 
\begin{equation}
    F(r,\theta,\phi) = \sum_{j=0}^{\infty}\frac{F_{j}(r,\theta,x) 
    }{r^{n_{F}+j}}
    \label{eq:formal},
\end{equation}
where the $r$-dependence of the $F_{j}$ is to be understood to be 
purely oscillatory. 
In spherical 
coordinates we expect for the algebraic dependence on $r$ 
that $\delta_{00}$, $\delta_{01}$ and 
$\delta_{11}$ are of order $1/r$, that $\delta_{02}$, $\delta_{03}$, 
$\delta_{12}$ and $\delta_{13}$ are of order $r^{0}$, and that 
$\delta_{22}$, $\delta_{23}$ and $\delta_{33}$ are of order $r$. We 
note that this ansatz for $\delta_{03}$ allows for a so-called 
Newman-Unti-Tamburini parameter which corresponds to a magnetic monopole 
in electrodynamics. It is unclear whether such terms have to be 
expected in the present context, and the used methods are not suited 
to answer this question. 

To 
fix the gauge freedom we consider coordinate transformations of the 
form below which do not change $\eta_{AB}$ to leading order.  
The wanted gauge transformations 
can be put into the form
\begin{eqnarray}
    \tilde{t}& = & t+\frac{T(r,\theta,x)+\alpha x}{r},
    \nonumber  \\
    \tilde{r} & = & r+\beta+\frac{R(r,\theta,x)}{r},
    \nonumber  \\
    \tilde{\theta}& = & \theta +\frac{Q(r,\theta,x)+\gamma x}{r^{2}},
    \nonumber  \\
    \tilde{\phi} & = & \phi+\frac{P(r,\theta,x)}{r^{2}}
    \label{eq:rinf1g},
\end{eqnarray}
where the $r$- and $x$-dependence of the potentials $T$, $R$, $Q$ and 
$P$ is understood as 
before to be purely oscillatory, and where $\alpha$, $\beta$ and $\gamma$ 
are functions of $\theta$ only. 
This implies for the transformed Minkowski 
metric in leading order 
\begin{eqnarray}
    \tilde{g}_{00} & = & -1+\frac{2\Omega}{r}(T_{,x}+\alpha),\quad \tilde{g}_{01}= 
    -\frac{1}{r}(T_{,r}+\Omega R_{,x}), \quad 
    \tilde{g}_{02}  =  -\Omega (Q_{,x}+\gamma),
    \nonumber  \\
    \tilde{g}_{03} & = & -\Omega \sin^{2}\theta P_{,x}, \quad  \tilde{g}_{11} = 
    1+\frac{2}{r}R_{,r}, \quad \tilde{g}_{12} = Q_{,r},
     \label{eq:rinf2g}\\
    \tilde{g}_{13} & = & \sin^{2}\theta P_{,r},\quad \tilde{g}_{22}=r^{2}\left(
    1+\frac{2\beta}{r}\right), \quad 
    \tilde{g}_{23}= Q_{,x}+\sin^{2}\theta P_{,\theta}, \quad \tilde{g}_{33}= 
    r^{2}\sin^{2}\theta\left(
    1+\frac{2\beta}{r}\right)
    \nonumber .  
\end{eqnarray}
For metrics obtained by solving the linearized Einstein equations on 
a Minkowski background, this implies that the functions $T$ to $P$ 
in (\ref{eq:rinf2g}) can be used to establish a certain gauge. 
By an appropriate choice of $P$, we can choose $\tilde{g}_{23}$ to 
vanish under the made assumptions. By considering higher orders of this 
expansion, this should be 
possible to all orders. Similarly by choosing $R$ and $\beta$
we obtain 
$\tilde{g}_{22}=r^{2}\tilde{g}_{11}$. The additional freedom can be used to have a 
vanishing $\delta_{00}$ and $g_{02}$. We assume that the gauge 
potentials $T$, $R$, $Q$ and $P$ are 
of the form $T=\sum_{m\in \mathbb{Z}}^{}(T_{m}^{+} (\theta)
e^{im(x+ \Omega r)}+T_{m}^{-} (\theta)
e^{im(x- \Omega r)})$. Therefore we had to add the functions $\alpha$, $\beta$ 
and $\gamma$ depending only on $\theta$
to make sure that terms in the $\delta_{AB}$ which are 
constant with respect to $x$ can be compensated. 
This would lead, however, to terms proportional to $x$ in 
$\delta_{00}$ and $\delta_{02}$.  Therefore we allow for a purely 
$\theta$-dependent $\delta_{00}$ and $\delta_{02}$ which will not 
enter the linearized field equations to retain periodic potentials in 
$x$ and $r$. Dropping the tilde, we thus choose the gauge (which is 
not exactly equivalent one used in the previous section, but close to it)
\begin{equation}
    g_{00}=-\left(1+\frac{f_{0}(\theta)}{r}\right), \quad g_{01} = \frac{c}{r},
    \quad g_{03} = a, \quad g_{33}=r^{2}\sin^{2}\theta 
    \left(1+\frac{F}{r}\right)
    \label{eq:rinf1},
\end{equation}
and 
\begin{equation}
    g_{02}=h_{0}(\theta),\quad
    g_{11}=\frac{1}{r^{2}}g_{22}=1+\frac{A}{r},\quad g_{12} 
    =\frac{\Theta}{r},\quad g_{13}=\Phi
    \label{eq:rinf2}.
\end{equation}
The gauge is fixed up to a free function of $r,\phi$ in $a$, 
and functions of $\theta$  only in $c$,  $\Theta$, and $\Phi$, 
since we assume periodicity in $r$ and $x$ of the terms 
$F_{n}(r,\theta,x)$ in (\ref{eq:formal}). 
There could be a contribution in order $r^{0}$ to $\Theta$, but this 
must be a function of $\theta$ alone as a consequence of the Einstein 
equations below and the periodicity condition in $r$.  We put this function 
equal to zero here to fix a gauge freedom.
This implies in leading order for the inverse metric
\begin{equation}
    g^{00}=-\left(1-\frac{f_{0}}{r}\right), \quad g^{01}=\frac{c}{r},
    \quad g^{03}=\frac{a}{r^{2}\sin^{2}\theta},
    \quad g^{33}=\frac{1}{r^{2}\sin^{2}\theta}
	\left(1-\frac{F}{r}\right)
    \label{eq:rinf1a},
\end{equation}
and 
\begin{equation}
    g^{02}=\frac{h_{0}}{r^{2}}, \quad g^{11}=r^{2}g^{22}=1-\frac{A}{r},\quad g^{13}=-\frac{\Phi}{
    r^{2}\sin^{2}\theta},\quad g^{12}= -\frac{\Theta}{r^{3}} .
    \label{eq:rinf2a}
\end{equation}

We obtain\\
\begin{proposition}\label{pro4.1}
    The linearized Einstein equations on a Minkowski background for 
    the metric (\ref{eq:rinf1}), (\ref{eq:rinf2}) lead  to two 
    wave equations for the functions $A$ and $a$,
    \begin{equation}
        A_{,rr}-\Omega^{2}A_{,\phi\phi}=0, \quad 
        a_{,rr}-\Omega^{2}a_{,\phi\phi}=0
        \label{eq:wave}.
    \end{equation}
    The remaining metric potentials follow in terms of quadratures.
\end{proposition}

Proof:\\
With relation
\begin{equation}
    R_{abcd}=\frac{1}{2}(g_{ad,bc}+g_{bc,ad}-g_{ac,bd}-g_{bd,ac})
    \label{eq:riemanne2}
\end{equation}
for the linearized
Riemann tensor we get for the Ricci tensor in lowest order in $1/r$
\begin{eqnarray}
    2rR_{00} & = & -2\Omega 
    c_{,r\phi}-2\Omega^{2}A_{,\phi\phi} -\Omega^{2}F_{,\phi\phi}
    \nonumber  \\
    2rR_{01} & = & \Omega(A+F)_{,r\phi}
    \nonumber  \\
    2rR_{02} & = & c_{,r\theta}-\Omega \Theta_{,r\phi}+\Omega(
    A+F)_{,\phi\theta}
    \nonumber  \\
    2R_{03} & = & -\Omega\Phi_{,r\phi}-a_{,rr}
    \nonumber  \\
    2rR_{11} & = & 2\Omega c_{,r\phi} +\Omega^{2}A_{,\phi\phi} 
    -(A+F)_{,rr}
    \nonumber  \\
    2rR_{12} & = & \Omega c_{,\theta\phi}-F_{,r\theta}+\Omega^{2} 
    \Theta_{,\phi\phi}
    \nonumber  \\
    2R_{13} & = & \Omega a_{,r\phi}
    +\Omega^{2}\Phi_{,\phi\phi}
    \nonumber  \\
    \frac{2}{r}R_{22} & = & 
    \Omega^{2}A_{,\phi\phi}-A_{,rr}
    \nonumber  \\
    2R_{23} & = & \Omega 
    a_{,\theta\phi}+\Phi_{,r\theta} 
    \nonumber  \\
    \frac{2}{r\sin^{2}\theta} R_{33} & = & \Omega^{2}F_{,\phi\phi}-F_{,rr}
    \label{eq:rinf3}.
\end{eqnarray}

It is a consequence of the equations for $R_{03}$ and $R_{13}$ that 
\begin{equation}
    \Omega \Phi_{,\phi}+a_{,r} = G_{1}(\theta)
    \label{eq:rinf4},
\end{equation}
where $G_{1}$ is a free function of $\theta$ only which is gauge 
invariant under transformations of the form (\ref{eq:rinf1g}). The equation for 
$R_{23}$ then implies 
\begin{equation}
    \Omega^{2} a_{,\phi\phi}-a_{,rr}=G_{2}(r,\phi),
    \label{eq:rinf5}
\end{equation}
where $G_{2}$ is a free function of $r$ and $\phi$ which  
reflects a gauge freedom and can be put equal to zero. Equation (\ref{eq:rinf5})
represents the first of the two wave equations.
We write the solution  in the form of a Fourier series
\begin{equation}
    a = \sum_{m\in \mathbb{Z}}^{}e^{im\phi}(a_{m}^{+}(\theta) 
    e^{i\Omega mr}+a_{m}^{-}(\theta)e^{-i\Omega mr})
    \label{eq:rinf6}.
\end{equation}
The reality condition for $a$ implies $a_{-m}^{\pm}=\bar{a}_{m}^{\pm}$.
Thus we get for $\Phi$
\begin{equation}
    \Phi = -\sum_{m\in \mathbb{Z}}^{}e^{im\phi}(a_{m}^{+}(\theta) 
    e^{i\Omega mr}-a_{m}^{-}(\theta)e^{-i\Omega mr})+G_{0}(\theta)
    \label{eq:rinf6a}.
\end{equation}

The equation for $R_{01}$ implies 
\begin{equation}
    A+F = G_{3}(\theta,\phi)+G_{6}(\theta,r)
    \label{eq:rinf7}.
\end{equation}
It is then a consequence of $R_{22}$ ($R_{33}$ is identically 
satisfied) that 
\begin{equation}
    A = \sum_{m\in \mathbb{Z}}^{}e^{im\phi}(A_{m}^{+}(\theta) 
    e^{i\Omega mr}+A_{m}^{-}(\theta)e^{-i\Omega mr})
    \label{eq:rinf8}.
\end{equation}
This gives the second wave equation. Again reality of $A$ 
implies $A_{-m}^{\pm}=\bar{A}_{m}^{\pm}$.

Equations $R_{00}$ and $R_{11}$ lead to 
\begin{equation}
    c = -\frac{\Omega}{2}\int_{r_{0}}^{r}A_{,\phi}dr +G_{4}(\theta)
    \label{eq:rinf9}.
\end{equation}
These equations also determine that the right-hand side of 
(\ref{eq:rinf7}) is only a function of 
$\theta$ if the periodicity in $r$ and $\phi$ is taken into account. 
Equations $R_{02}$ and $R_{12}$ then imply 
\begin{equation}
    \Theta 
    =-\frac{1}{2}\int_{r_{0}}^{r}A_{,\theta}dr +G_{5}(\theta)
    \label{eq:rinf11a}.
\end{equation}
This completes the proof. 

\begin{remark}\label{4.1}
    If the black holes have equal `mass', i.e.\ equal combination of mass 
    and angular momentum which can be defined via the Komar 
    integral below, the spacetime has an additional discrete symmetry, it 
    is invariant in a suitably defined coordinate system under the
    transformation $\phi\to-\phi$. This implies for (\ref{eq:rinf6}) and 
    (\ref{eq:rinf8}) $A_{m}^{+}=\bar{A}_{m}^{-}$ and $a_{m}^{+}=\bar{a}_{m}^{-}$. 
    In this case no additional boundary conditions at infinity need to 
    be given. A Sommerfeld condition which is typically 
    considered at finite radius would only allow trivial solutions in 
    this example if imposed at infinity. 
\end{remark}

\begin{remark}\label{4.2}
    In case the functions $a$ and $\Phi$ have leading terms  of order 
    $1/r$, i.e.\ if there is no NUT-parameter, the following equations for 
    the Ricci tensor (\ref{eq:rinf3}) change
    \begin{eqnarray}
         2rR_{03} & = & c_{,r\phi}-\Omega\Phi_{,r\phi}-a_{,rr}+2\Omega 
         A_{,\phi\phi},
        \nonumber  \\
        2rR_{13} & = & \Omega a_{,r\phi}+\Omega c_{,\phi\phi}-A_{,r\phi}
        +\Omega^{2}\Phi_{,\phi\phi},
        \nonumber  \\
        2rR_{23} & = & \Omega 
        a_{,\theta\phi}+\Phi_{,r\theta} 
        +\Theta_{,r\phi}-A_{,\theta\phi}
        \label{eq:rinf3x}.
    \end{eqnarray}
    These equations again imply wave equation (\ref{eq:rinf5}) for $a$ and 
    \begin{equation}
        \Phi = -\frac{1}{\Omega}\int_{\phi_{0}}^{\phi}a_{,r}d\phi 
        +\frac{3}{2}\int_{r_{0}}^{r}A_{,\phi}dr +G_{7}(\theta,r)
        \label{eq:phin}.
    \end{equation}
\end{remark}

The ansatz (\ref{eq:formal}) already implies that the ADM mass and 
additional asymptotic multipoles cannot be defined due to the oscillatory 
behavior of the metric functions. We will show that it is also not 
possible to use the Komar integral asymptotically in a standard way 
to define a conserved quantity. 
This  integral  can be used to 
relate a locally calculated mass to the ADM mass for an 
asymptotically flat spacetime with a stationary Killing vector. The 
idea is to evaluate a surface integral at finite radius $R$ and then to 
take the limit $R\to \infty$. 
Basically one uses that $\star d\xi$ is an exact differential which 
means that one can apply Gauss' theorem. We get for an integration 
over a sphere with $t=const$, $r=const$
\begin{equation}
    \int_{S}^{}\xi_{[A,B]}g^{0A}g^{1B}\sqrt{-g}d\theta d\phi=0
    \label{eq:komar1}.
\end{equation}

To calculate the integral near the horizon we need 
the inverse of the 4-dimensional 
metric 
\begin{equation}
    g^{00}=-\frac{1}{f}+fk_{a}k^{a},\quad g^{0a}=-fk^{a},\quad 
    g^{ab} = fh^{ab}
    \label{eq:komar2},
\end{equation}
where spatial indices are raised and lowered with $h_{ab}$. In the 
quasi-isotropic gauge we  get with the results of section 3 for the 
surface integral that only the term 
\begin{equation}
    \int_{S}^{}\sqrt{-g}d\theta d\phi g^{00}g^{11} \xi_{[0,1]}
    \label{eq:komarh}
\end{equation} contributes.  This leads with $\mathcal{A}_{0}=\kappa 
f_{0}$ to 
\begin{equation}
    \frac{1}{2}\int_{S}^{}\sin\theta d\theta d\phi 
    \frac{f_{,r}}{\mathcal{A}}= \frac{4\pi}{\kappa}
    \label{eq:komarh2}.
\end{equation}
For  a single black hole the constant $\kappa=1/(2m)$. 
The Komar integral is of course only defined up to a scaling of the 
helical Killing vector, see the remarks in \cite{friedman}. 

To check whether the surface integral can be defined for $r\to \infty$, we 
determine the integral for $r=r_{0}$ where $r_{0}\gg R$ and study 
whether the limit $r_{0}\to\infty$ exists. This could be possible if one 
uses the periodicity of the functions in $\phi$ in the 
$\phi$-integration over a complete period at a finite value of $r$. 
As we will show below, this will not be the case because of 
the bilinear terms in the integrand. Note that the Killing vector 
reads in the used coordinates 
$\xi=\partial_{t'}+\Omega\partial_{\phi'}$. 
It is readily seen that the integral can only exist if $G_{1}=0$, 
since the corresponding terms in the integrand are of order $r^{2}$. 
If we assume that this is the case, we get for the surface integral
\begin{eqnarray}
    \frac{1}{2}\int_{S}^{}r\sin\theta 
    d\theta d\phi (
    \Omega^{2}\Phi F_{,\phi}+a\Omega(2+F_{,r}))
    \label{eq:komar3}.
\end{eqnarray}
The integral can only exist if these terms vanish after integration 
with respect to $\phi$ since the integrand diverges as $r$. Writing 
the integrands as a Fourier series as we have done in the proof of proposition 
4.1, we get after integration 
with respect to $\phi$ that the integrand is proportional to 
\begin{equation}
    a_{0}^{+}+a_{0}^{-}-i\Omega \sum_{m\in 
    \mathbb{Z}}^{}m(a_{m}^{+}\bar{A}_{m}^{-}e^{2i\Omega mr}-
    a_{m}^{-}\bar{A}_{m}^{+}e^{-2i\Omega mr}) 
    \label{eq:komar3a}.
\end{equation}
This expression can only vanish if 
$a_{m}^{+}\bar{A}_{m}^{-}+\bar{a}_{m}^{-}A_{m}^{+}=0$ for $m>0$. In 
the equal mass case, this is only possible if either $a_{m}$ or $A_{m}$ 
vanish. If  $a$ and $\Phi$ have leading contributions in order $1/r$ 
as in remark ~\ref{4.2}, the terms of order $0(r^{0})$in the surface integral  are due to 
(\ref{eq:komarh}) and are of the form 
\begin{equation}
    \sum_{m\in 
	\mathbb{Z}}^{}m(A_{m}^{-}\bar{A}_{m}^{+}e^{2i\Omega mr}-
	A_{m}^{+}\bar{A}_{m}^{-}e^{-2i\Omega mr}) 
    \label{eq:komarnut}.
\end{equation}
The integral can only exist if the terms $A_{m}^{+}A_{m}^{-}$ vanish for all 
$m\neq 0$. Since there cannot be purely `outgoing' or `ingoing' waves 
in the case of a helical Killing vector, this condition 
will lead to the axisymmetric 
case. Thus the surface integral cannot be defined asymptotically in the 
presence of a helical Killing vector unless there is in addition an 
asymptotically axial Killing vector. In this case the integral 
just gives the expected value $M+\Omega J$, the combination of mass 
and angular momentum corresponding to a helical Killing vector. 

Gibbons and Stewart \cite{gibstew} showed that periodic boundary 
conditions are incompatible with a smooth null infinity. This is in 
accordance with the ansatz (\ref{eq:formal}) as can be seen from the 
following consideration: We define the standard null-tetrad of Minkowski 
spacetime, 
\begin{equation}
    k^{a}= \frac{1}{\sqrt{2}}\left(
    \partial_{t'}+\partial_{r}\right),\quad 
    m^{a}= \frac{1}{\sqrt{2}}\left(
    -\partial_{t'}+\partial_{r}\right),\quad
	t^{a}= \frac{1}{\sqrt{2}r}\left(\partial_{\theta}+\frac{i}{\sin\theta}
	\partial_{\phi'}\right)
    \label{eq:riemann17},
\end{equation}
We define the Weyl scalars as in \cite{stephani} (we can use the above 
tetrad since we are only considering a linearization on a Minkowski 
background)
\begin{eqnarray}
    C_{1} & = & 2C_{abcd}m^{a}m^{c}t^{b}t^{d} 
    \nonumber,  \\
    C_{2} & = & -C_{abcd}m^{a}t^{b}(k^{c}m^{d}+\bar{t}^{c}t^{d}) 
    \nonumber,  \\
    C_{3} & = & 2C_{abcd}m^{a}t^{b}k^{c}\bar{t}^{d}
    \nonumber,  \\
    C_{4} & = & -C_{abcd}k^{a}\bar{t}^{b}(k^{c}m^{d}+\bar{t}^{c}t^{d}) 
    \nonumber,  \\
    C_{5} & = & 2C_{abcd}k^{a}k^{c}\bar{t}^{b}\bar{t}^{d}
    \label{eq:riemann18}.
\end{eqnarray}

Determining the components of the Riemann tensor for the asymptotic metric of 
proposition 4.1, we get for the Weyl scalars in leading order 
\begin{eqnarray}
    C_{1} & = & -\frac{1}{2r}(A_{rr}+2\Omega 
    A_{r\phi}+\Omega^{2}A_{\phi\phi}),
    \nonumber    \\
    C_{2} & = & C_{3}= C_{4} = 0, \nonumber \\
    C_{5} & = & -\frac{1}{2r}(A_{rr}-2\Omega 
    A_{r\phi}+\Omega^{2}A_{\phi\phi}).
    \label{eq:rinf12}
\end{eqnarray}
Thus the Petrov type is N. The Weyl scalars vanish for $r\to\infty$, 
but this limit is in general not defined for $rC_{i}$, $i=1,\ldots,5$ 
because of the oscillatory behavior of the metric functions. Thus in 
accordance with \cite{gibstew}, there is no smooth $\mathcal{I}$ and 
no peeling in this case even if we assume that the metric tends asymptotically to 
the Minkowski metric.

\section{Outlook}
In the previous sections  we have given a set of equations describing binary 
black hole spacetimes with a helical Killing vector. The equations 
have regular singularities at the Killing horizons and the light 
cylinder, and a non-regular singularity at infinity. This leads to a 
set of five equations which could be useful for a numerical 
implementation. The equations appear to be well suited for the 
multi-domain spectral method used in \cite{ggb2}, see also 
\cite{lorene}. It is straight forward to include regular 
singularities in the spectral formalism in the adapted coordinates we used for 
the analytical discussion, since the formal expansions we were 
discussing is very close to the philosophy of a spectral expansion. 
The main difficulty from a numerical point of view seems to be the 
oscillatory behavior at infinity. Typically a cut-off at some finite 
radius is used, but it is unclear which boundary conditions have to be 
used there. A possibility would be to match the solution at some 
large radius to an analytical solution of the linearized Einstein 
equations where it is not yet clear on which background the equations 
can be linearized (the asymptotic form of the solutions is still an 
open question). The main problem will be in any case the numerical resolution of the 
oscillatory metric close to infinity. 

From a mathematical point of view the most interesting question is 
whether there exist solutions with two regular Killing horizons in a 
vacuum spacetime with a helical Killing symmetry. In this paper we have 
only considered formal expansions of the 
metric in the vicinity of the singularities. The fact that the 
solution close to the horizons contains two free functions of the 
angular variables gives hope that such solutions might exist 
globally, but this needs to be proven. In case such solutions exist,  it would be interesting 
to obtain the precise asymptotic behavior, whether the metric tends 
to the Minkowski metric asymptotically, and whether a NUT parameter 
is needed. Numerical results could give hints 
on how to answer these mathematical questions.

The physical relevance of the studied model is clearly to obtain
fully relativistic values for the ISCO and to get initial data for 
numerical calculations of the last phase of the binary system. In a 
real physical situation, the helical symmetry will be only an 
approximate symmetry. Therefore it would be interesting to study 
perturbations of a spacetime with an exact helical Killing vector 
studied here. There have been activities in this direction: in \cite{detweiler2}, the Killing symmetry holds only in 
a finite region of space and time, the spacetime is asymptotically 
matched to a wave-zone. An approximate Killing vector was 
considered in \cite{cook}. 

\textbf{Acknowledgement:}\\
I thank R.~Beig, L.~Blanchet, S.~Bonazzola, B.~Carter, P.~Forgacs, 
J.~Frauendiener, P.~Grandcl\'ement, 
E.~Gourgoulhon, who interested me in the subject, D.~Maison, J.~Novak, 
and B.~Schmidt for helpful discussions and hints. In addition I thank 
the unknown referee for clarifications, corrections and useful comments. This work was 
supported in part by the Marie-Curie program of the European Union 
and the Schloessmann foundation.

\end{document}